# Functional Programming and Streams

John MacCormick, Dickinson College, February 2023.

Abstract: This document is intended as a stand-alone textbook chapter to be used for introducing some functional programming concepts into a course in which the primary teaching language is Java. For details of the approach, please see the paper "Functional programming, in the data structures course, in Java" by J. MacCormick, which appears in the *Journal of Computing Sciences in Colleges* (2023).

## 1   Some background on functional programming

The design and implementation of programming languages is a large and important subfield of computer science. In this chapter, we examine *functional programming*, which is one of the most important ideas within the theory of programming languages.  Programming languages are often described as belonging to various categories or *paradigms*. Examples of these programming language paradigms include *imperative* languages, *functional* languages, and *logic programming* languages. Most modern languages include features from multiple paradigms. For example, Java and Python were designed primarily as imperative languages, but they include many aspects of functional programming. Examples of languages that were designed as functional languages include Lisp, F#, and Haskell.

In this chapter, we first examine one of the most fundamental ideas in functional programming: *lambda expressions*. Then, we see how lambda expressions can be used in Java's Stream API to process data sets efficiently and elegantly.

### 1.1   A note on the use of Java and Python in this chapter

Although Java and Python are not particularly good examples of functional programming languages, the examples in this chapter use only Java and Python. It would take us too far afield to study a more purely functional language. If some of the Java examples seem a little awkward, keep in mind that we are deliberately adopting a compromise between understanding the ideas of functional programming and exploiting our existing familiarity with Java. If you are not familiar with Python, do not be concerned. It is easier to demonstrate some of functional programming ideas in Python, compared to Java. Therefore, we use some initial examples from Python. But it will be sufficient to understand the Java examples without having a detailed understanding of the Python examples.

## 2   Lambda expressions

In the rest of this chapter, the word *function* usually refers to a subroutine in a programming language that can accept parameters and return values. Different programming languages refer to functions using different terminology. In Java, for example, functions are called *methods*. In Python, functions are simply called functions.

A key aspect of functional programming languages is that they treat functions the same as other data types. This is often described using the phrases "functions are first-class citizens" or "functions are first-class objects." The most important consequence of this first-class status is that in a functional language, a function can be a parameter in another function.

## 2.1 Functions as parameters in Python

It is particularly easy to demonstrate this in Python:

```python
def add5(x):
    return x + 5

def multBy3IfPositive(x):
    if x > 0:
        return 3 * x
    else:
        return 0

def applyToSeven(f):
    return f(7)

def applyToMinusNine(f):
    return f(-9)
```

Here, the functions `applyToSeven` and `applyToMinusNine` both accept a single parameter which is expected to be a function. For example, we can use `add5` as the parameter for `applyToSeven`:

```
>>> applyToSeven(add5)
12
```

And of course, we can get a different result if we send in a different function as the parameter:

```
>>> applyToSeven(multBy3IfPositive)
21
```

> Example problem 1.
>
> Check your understanding by working out the results of the following two function calls:
>
> >>> applyToMinusNine(add5)
>
> >>> applyToMinusNine(multBy3IfPositive)
>
> Solutions to all example problems are given at the end of this chapter.

## 2.2 Functions as parameters in Java

In Java, it is not quite so easy to pass a function as a parameter, when compared to Python. The Java code below is the simplest equivalent of the above Python code. As you will see, it is comparatively ugly. It relies on Java interfaces to simulate the ability to pass a function as a parameter. In this example, we import the `Function<T, R>` interface from `java.util.function`, which is the Java package that supports functional programming. The `Function<T, R>` interface represents a function that accepts a single parameter of type `T` and returns a value of type `R`. Because the `add5` function accepts a single integer parameter and returns an integer, we can create the effect of functional programming with `add5` by creating an `Add5` class that implements the `Function<Integer, Integer>` interface. In the `main()` method below, we create an instance of the `Add5` class. The name of that instance is `add5`, and we can pass this instance, which truly is a Java object, as a parameter to the `applyToSeven()` and `applyToMinusNine()` methods.

```java
import java.util.function.Function;

public class FunctionParameterDemo {

    public static class Add5 implements Function<Integer, Integer> {
        public Integer apply(Integer x) {
            return x + 5;
        }
    }

    public static class MultBy3IfPositive
                implements Function<Integer, Integer> {
        public Integer apply(Integer x) {
            if (x > 0) {
                return 3 * x;
            } else {
                return 0;
            }
        }
    }

    public static Integer applyToSeven(Function<Integer, Integer> f) {
        return f.apply(7);
    }

    public static Integer applyToMinusNine(Function<Integer, Integer> f) {
        return f.apply(-9);
    }

    public static void main(String[] args) {
        Add5 add5 = new Add5();
        MultBy3IfPositive multBy3IfPositive = new MultBy3IfPositive();
        int val1 = applyToSeven(add5); // val1 = 12
        int val2 = applyToSeven(multBy3IfPositive); // val2 = 21
    }
}
```

Example problem 2.

Determine the value of the following two method calls, assuming they were inserted at the end of the above `main()` method:

```
applyToMinusNine(add5)
applyToMinusNine(multBy3IfPositive)
```

## 2.3 Functional interfaces and lambda expressions in Java

Notice that the interface implemented by `add5` and `multBy3IfPositive` has the following special property: it has only a single abstract method. The name of this method is `apply()`. In Java, an interface with exactly one abstract method is called a *functional interface*. As we will see later, functional interfaces are important because they can be implemented efficiently and elegantly using *lambda expressions*. Lambda expressions are a shorthand notation for functional interfaces that allow us to avoid using the ugly style of Java functional programming above. To see lambda expressions in action, we will first return to Python.

## 2.4 Named and anonymous values

In any computer program written in any programming language, some of the values will typically be *named* whereas other values will be *anonymous*. For example, here are two different ways of printing out the square root of 5 in Python:

```
print(math.sqrt(5)) # anonymous

x = 5
print(math.sqrt(x)) # named
```

In the first call to `sqrt()`, the value 5 is anonymous because it is not referred to using the name of a variable. In the second call to `sqrt()`, the value 5 is referred to using the name `x`.

## 2.5 The lambda keyword in Python

Recall that in functional programming, we can use functions as parameters. Therefore, it should be possible to use the same two techniques when sending a function as a parameter: we can use either a named function, or an anonymous function. We have already seen how to do this using named functions. One of the examples from earlier is repeated here for concreteness:

```
>>> applyToSeven(add5)
```

This sends the named function `add5` as a parameter to an invocation of the function `applyToSeven()`. But how can we do this using an anonymous function? In Python, we can create an anonymous function using the keyword `lambda`, as in the following example.

```
>>> applyToSeven(lambda x: x+5)
```

So, a lambda expression is just a way of describing a function without giving it a name. The Python snippet "`lambda x: x+5`" means "the function that takes a parameter `x` and returns `x+5`." It would perhaps be less confusing if `lambda` were called something else like `anonFunction`, but there are good historical and theoretical reasons for adopting the keyword `lambda`. "Lambda" is the name of the Greek letter $\lambda$. In the 1930s, the mathematician Alonzo Church used the Greek letter $\lambda$ as the main symbol in describing a framework for computing with functions—a framework that we now call the *lambda calculus*. The lambda calculus lies at the heart of functional programming and the theory of computation, but it would take us too far afield to study that connection here.

To summarize: in Python, a lambda expression defines an anonymous function. There is never any need to be confused by lambda expressions. If you see some code that contains a confusing lambda expression, you can simply rewrite it by first creating a named function that performs the effect of the lambda expression, then substituting the new name for the lambda expression. For example, the Python snippet

```
performStrangeAction("nonsense", \
      lambda apple, banana, grape: apple + banana.gimble(2*grape))
```

can be rewritten as

```
def weirdFunction(apple, banana, grape):
    return apple + banana.gimble(2*grape)

performStrangeAction("nonsense", weirdFunction)
```

The above example also shows how lambda expressions in Python can describe functions with multiple parameters.

> Example problem 3.
>
> Assuming the definitions given earlier, what is the output of the following snippets of Python?
>
> ```
> applyToSeven(lambda potato: potato%4 + potato*potato)
> applyToMinusNine(lambda oak: math.factorial(oak+12) * oak)
> ```

## 2.6   Lambda expressions in Java

In Java, lambda expressions do not use the keyword `lambda`. Instead, they employ the arrow notation "`->`" immediately after the function parameters. For example, the equivalent of the Python snippet "`lambda x: x+5`" in Java is "`x -> x+5`". In both cases, the meaning of the lambda expression is approximately "the function that receives a parameter x and returns x+5." The true meaning of a Java lambda expression is more complex, but we won't describe the details here. It is enough to know that the Java lambda expression uses the functional interfaces mentioned earlier, creating an anonymous class and an anonymous instance so that the desired functional programming effect is achieved.

We can rewrite our earlier Java examples using lambda expressions. The declarations of `applyToSeven()` and `applyToMinusNine()` remain the same, but they are repeated here for convenience:

```
    public static Integer applyToSeven(Function<Integer, Integer> f) {
        return f.apply(7);
    }

    public static Integer applyToMinusNine(Function<Integer, Integer> f) {
        return f.apply(-9);
    }
```

The rest of the code is much more compact because there is no need to declare any classes that implement functional interfaces. Our main method can instead be

```java
public static void main(String[] args) {
    int val3 = applyToSeven(x -> x + 5); // val3 = 12
    int val4 = applyToSeven(x -> {
        if (x > 0) {
            return 3 * x;
        } else {
            return 0;
        }
    }); // val4 = 21
}
```

Note the use of a lambda expression that spans multiple lines and contains multiple blocks of code. In practice, lambda expressions are usually kept short and simple, but they can be as complex as desired.

Example problem 4.

Determine the values of the following two method calls, assuming they were inserted at the end of the above `main()` method:

```java
int val5 = applyToMinusNine(x -> (x + 1) * (x + 2));
int val6 = applyToMinusNine(z -> {
    if (z > 10) {
        return 100;
    } else if (z < -100) {
        return -100;
    } else {
        return z * 10;
    }
});
```

Lambda expressions in Java can also have multiple parameters. This is achieved by listing the parameters inside parentheses before the "->" symbol. For example, the lambda expression `(x, y) -> x*x + 2*y` could represent a method `f()` such as the following.

```java
public int f(int x, int y) {
    return x*x + 2*y;
}
```

Example problem 5.

Write the mathematical function $g(u, v, w) = \sqrt{u^2 + v^2 + w^2}$ as a Java lambda expression.

# 3 The Java Stream API

In Java, `Stream<T>` is an interface for performing operations on sequences of objects of type `T`. There are also specialized streams for performing operations on sequences of some primitive types: `IntStream` for `int` values, `LongStream` for `long` values, and `DoubleStream` for `double` values.

Common operations used on `Stream`s include `count()`, `filter()`, `map()`, `mapToInt()`, `foreach()`, and `reduce()`. `IntStream`, `LongStream`, and `DoubleStream` also have the operation `sum()`. We study only these seven operations. The Java API documentation lists other available operations.

Each operation is defined as either an *intermediate operation* or a *terminal operation*. The seven common operations that we will study are classified as follows.

- Intermediate operations: `filter()`, `map()`, `mapToInt()`.
- Terminal operations: `count()`, `foreach()`, `reduce()`, `sum()`.

To perform a computation on a `Stream`, we can apply zero or more intermediate operations in sequence, followed by a single terminal operation. For example, a computation might consist of the following sequence of operations: `filter()`, `filter()`, `map()`, `filter()`, `count()`.

## 3.1 Creating a Stream

There are numerous ways to create a `Stream` in Java. Several of these approaches are shown in the following code snippet.

```java
// Approach 1: Create directly from an array via Stream.of()
String[] array = { "bat", "cat", "bird", "mad", "catch", "ditch" };
Stream<String> stream1 = Stream.of(array);

// Approach 2: Create directly from multiple arguments via
// Stream.of()
Stream<String> stream2 = Stream.of("bat", "cat", "bird", "mad",
          "catch", "ditch");

// Approach 3: Convert any Java Collection using the collection's
// stream() method
List<String> list = Arrays.asList("bat", "cat", "bird", "mad",
          "catch", "ditch");
Stream<String> stream3 = list.stream();

// Approach 4: Stream from a file using Files.lines
Stream<String> stream4 = Files.lines(Paths.get("data/words.txt"));

// Approach 5: Use range() or rangeClosed()
IntStream range1 = IntStream.range(5, 10); // 5,6,7,8,9
IntStream range2 = IntStream.rangeClosed(5, 10); // 5,6,7,8,9,10
```

Example problem 6.

(i) Write some code that would create a stream containing the following sequence as `Double` objects: 23.4, 69.7, -25.88, 31.3363.

(ii) Repeat part (i), this time creating a stream containing primitive `double` values.

(iii) Write code creating a stream consisting of integers from 100 to 200 inclusive.

The next section introduces each of our seven common operations, while also gradually building our understanding of how to combine the operations into more interesting stream computations.

## 3.2 Stream operations

We now examine seven stream operations in detail.

### 3.2.1 count()

The `count()` operation returns the number of elements in a stream.

```
Stream<String> stream = Stream.of("bat", "cat", "bird");
long numElements = stream.count(); // numElements = 3
```

Example problem 7.

Write a snippet of code that uses the Stream API to count the number of lines in a file called "`GreatGatsby.txt`".

### 3.2.2 sum()

The `sum()` operation returns the sum of the elements in a stream. It can only be applied to the specialized numeric streams `IntStream`, `LongStream`, and `DoubleStream`, as in the following example.

```
DoubleStream stream = DoubleStream.of(1.5, 2.4, -0.1);
double total = stream.sum(); // total = 3.8
```

Example problem 8.

Write a snippet of code that uses the Stream API to add the integers from 27 to 159 inclusive. Hint: `IntStream.range()` or `IntStream.rangeClosed()` make this very easy.

### 3.2.3 filter()

The `filter()` operation applies a Boolean test function to every element in the input stream. Elements that fail the test (i.e. return `false`) are discarded, whereas elements that pass the test (i.e. return `true`) enter the output stream to be processed by the next operation in the computation. This is our first example of using lambda expressions with `Stream`s, because the Boolean test function can be described with a lambda expression. Consider the following example.

```
Stream<String> stream = Stream.of("bat", "cat", "bird", "mad",
                                  "catch", "ditch");
Stream<String> newStream = stream.filter(word -> word.startsWith("ca"));
```

Here, `stream` is the input to the computation. It is a `Stream<String>` containing the elements "bat", "cat", "bird", "mad", "catch", "ditch". The output of the computation is `newStream`, which also has the datatype `Stream<String>`. But `newStream` contains only the elements beginning with "ca", so it consists of the two elements "cat" and "catch".

If the lambda expression in the snippet above is confusing, remember that we can always rewrite lambda expressions using named functions. Let's do this now for the above lambda expression, `word -> word.startsWith("ca")`.

The first step is to determine what functional interface this lambda expression is an instance of. To do that, we consult the documentation of the `filter()` method in the `Stream` interface. In the documentation, the signature of the `filter()` method is given as

`Stream<T> filter(Predicate<? super T> predicate)`

Until we have more familiarity with streams and lambda expressions, it will be best to ignore the type wildcard ("?"). So let's think of the parameter as having datatype `Predicate<T>`. We again consult the official Java documentation, this time looking up `Predicate<T>`. As expected, it is a functional interface, which means it has exactly one abstract method. The signature of this method is `boolean test(T t)`. Note that our `filter()` method is being applied to a stream of strings, `Stream<String>`. So for this example, the type parameter `T` has value `String`. Therefore, to rewrite the lambda expression `word -> word.startsWith("ca")`, we need to implement the interface `Predicate<String>`. Specifically, we will need to implement the method `test(String t)` so that its parameter is called `word` and its return value is `word.startsWith("ca")`. Putting this together, we obtain the following code.

```java
class StartsWithCA implements Predicate<String> {
    public boolean test(String word) {
        return word.startsWith("ca");
    }
}
StartsWithCA startsWithCA = new StartsWithCA();
Stream<String> stream = Stream.of("bat", "cat", "bird", "mad",
                    "catch", "ditch");
Stream<String> newstream = stream.filter(startsWithCA);
```

Once we have obtained the new, filtered stream `newstream`, we can do further computations on it. For example, we can count the elements:

```java
Stream<String> newstream = stream.filter(word -> word.startsWith("ca"));
long numStartWithCA = newstream.count(); // numStartWithCA = 2
```

However, this is usually written in a more compact form. Instead of defining a new local variable for each stream created by the intermediate operations, we can immediately apply the next operation via a method call:

```
long numStartWithCA = stream
            .filter(word -> word.startsWith("ca"))
            .count(); // numStartWithCA = 2
```

This is equivalent to the previous snippet, but it is more compact and more readable once you get used to the syntax. Each operation in the computation is written as a method call beginning with "`.`" and placed on a new line for readability.

> Example problem 9.
>
> Building on your solution to the previous example problem, write a snippet of code that uses the Stream API to add the *odd* integers from 27 to 159 inclusive.

### 3.2.4 foreach()

The `foreach()` operation applies an action to every element in the input stream. It is a terminal operation and should only be used for producing output at the end of a computation. Never use the `foreach()` operation to process elements in an intermediate operation. For our purposes, the only use of `foreach()` is to print out the elements of a stream using methods such as `System.out.print()` and `System.out.println()`, as in the following example.

```
Stream<String> stream = Stream.of("apple", "banana", "bagel");
stream.forEach(word -> System.out.println(word));
```

This snippet produces the output

```
apple
banana
bagel
```

We have already discussed how it is always possible to rewrite a lambda expression as a named instance of a functional interface. There is yet another way to rewrite certain simple lambda expressions. If the lambda expression does nothing except invoke a method that already has a name, you can refer directly to the method using Java's *method reference* operator, "`::`". For example, the snippet `System.out::println` is equivalent to the lambda expression `x -> System.out.println(x)`. Method references are often used with the `foreach()` operation, as in the following example (which is equivalent to the previous example).

```
Stream<String> stream5 = Stream.of("apple", "banana", "bagel");
stream5.forEach(System.out::println);
```

> Example problem 10.
>
> Write a snippet of code that uses the Stream API to print the integers from 27 to 159 inclusive on separate lines.

### 3.2.5 map()

The `map()` operation is used to apply a method to each element of the input stream. In mathematical notation, the result of mapping the function $f$ onto the sequence $x_1, x_2, x_3, x_4, \ldots$ is $f(x_1), f(x_2), f(x_3), f(x_4), \ldots$ . The following is an example using Java streams.

```
Stream<String> stream = Stream.of("apple", "banana", "bagel");
stream
    .map(word -> word.toUpperCase() + "***")
    .forEach(System.out::print);
```

The output is

> APPLE***BANANA***BAGEL***

Example problem 11.

> Suppose a file `info.txt` stores a nonempty string on each line. Write a snippet of code that uses the Stream API to print the first character of each line in the file on a separate line.

### 3.2.6 mapToInt()

The `mapToInt()` operation is identical to `map()`, except that the mapped function must return an `int`, and therefore the resulting stream is an `IntStream`. For example, we can find the location of the letter "e" in each element of a `Stream<String>` as follows.

```
Stream<String> stream = Stream.of("apple", "banana", "bagel");
stream
    .mapToInt(word -> word.indexOf('e'))
    .forEach(System.out::println);
```

The output is

> 4
> -1
> 3

Example problem 12.

> Write a snippet of code that uses the Stream API to print the length each line in the file `info.txt` on a separate line.

### 3.2.7 reduce()

The `reduce()` operation is a terminal operation, which is used to combine all the elements of a stream into a single output. Perhaps this operation would be easier to understand if it were called "combine." But the term "reduce" also makes sense, because we are "reducing" an entire stream into a single output.

The `reduce()` operation is a little more elaborate than the others, so we introduce it by way of an example. For the moment, we abandon streams and go back to processing data in an ordinary array. Suppose we have an array of strings and we would like to take the first character from each string and combine these into a single string. For example, the array `{"apple", "banana", "bagel"}` will produce the output "abb". This can be achieved by the following Java code.

```java
        String[] array = { "apple", "banana", "bagel" };
        String initialValue = "";
        String resultSoFar = initialValue;
        for (String newElement: array) {
            resultSoFar = resultSoFar + newElement.charAt(0);
        }
```

The idea behind this algorithm is obvious. We move through the array accumulating any results in the variable `resultSoFar`. Each time we process a new element, we combine it with the existing results (by adding the first character of `newElement`, in this particular case), producing an updated value for `resultSoFar`. We also specify an initial value for the results, which is stored separately for clarity in the variable `initialValue`. When the algorithm terminates, the final result can be found in the accumulator variable `resultSoFar`.

We can refactor the above snippet so that the process of accumulating results is factored out into the separate method `accumulate()`, as follows.

```java
public static String accumulate(String resultSoFar, String newElement) {
      return resultSoFar + newElement.charAt(0);
}
public static void main(String[] args) throws IOException {
      String[] array2 = { "apple", "banana", "bagel" };
      String initialValue = "";
      String resultSoFar = initialValue;
      for (String newElement : array2) {
            resultSoFar = accumulate(resultSoFar, newElement);
      }
      System.out.println(resultSoFar);
}
```

Now we can describe the `reduce()` streaming operation using the vocabulary from the above example. The operation has two parameters, which correspond to the `initialValue` variable and the `accumulate()` method. Thus, the `reduce(initialValue, accumulate)` streaming operation applies the accumulate method successively to each element of the stream, using the given initial value for initialization. The accumulate method is usually specified by a lambda expression.

Returning to our concrete example, the following code shows how to produce a string consisting of the initial characters of each element in a stream.

```
Stream<String> stream = Stream.of("apple", "banana", "bagel");
String firstLetters =
    stream.reduce(
            "", // first parameter is the initial value, an empty String
            (resultSoFar, newElement) -> resultSoFar + newElement.charAt(0)
            // second parameter (above) is the 'accumulate' function,
            // written as a lambda expression with two parameters
    );
```

Note how the lambda expression representing the accumulator function has two parameters, just like the `accumulate()` method in the earlier example above:

`(resultSoFar, newElement) -> resultSoFar + newElement.charAt(0)`

Note that we have studied the *two-parameter* form of the `reduce()` operation. There is also a one-parameter form and a three-parameter form, but we do not study those here.

> Example problem 13.
>
> Suppose a file `numbers.txt` stores a number written in decimal notation on each line. Write a snippet of code that uses the Stream API to compute the sum of the numbers in the file using the `java.math.BigDecimal` class, which guarantees that no precision will be lost when dealing with decimal numbers. (It would be important to use this approach when performing a financial calculation, for example.)

## 3.3  Parallel streams

One advantage of the Java Stream API is that computations can be parallelized with essentially no effort. Every stream has an internal setting that determines whether it operates in a sequential fashion or a parallel fashion. If a stream is set to operate in parallel, the Java stream library will attempt to split the stream into chunks which are fed into separate threads running simultaneously. Certain computations benefit greatly from this parallelism and will complete more quickly. Indeed, if there are $N$ CPU cores available then the computation could in principle be sped up by a factor of $N$ (or even more if we take hyperthreading into account). On the other hand, some computations cannot benefit from parallelization. When operating in parallel mode, such a computation may in fact run more slowly than its sequential version, due to the overhead of setting up the parallel streams.

Any stream can be converted to a parallel stream by applying the `parallel()` operation to it. A stream can be converted to sequential mode by applying the `sequential()` operation. The following is an example of code that benefits greatly from employing parallelism when the array `values[]` is very large. We assume a Boolean method `isPrime()` is available, which returns `true` if its argument is a prime number.

```
int[] values = …
numPrimes = IntStream.of(values)
                .parallel()
                .filter(n -> isPrime(n))
                .count();
```

To perform the same computation sequentially, replace the "`parallel()`" with "`sequential()`". The program `ParallelPrimesDemo.java` (see listing below) compares the two approaches empirically.

> Example problem 14.
>
> Run the program `ParallelPrimesDemo.java` (see listing below) on your own computer and determine the average speed-up factor when the computation is done in parallel. How does this compare to the number of CPU cores and/or hyperthreads available on your computer?

### 3.4 Program listing of ParallelPrimesDemo.java

```java
import java.util.Random;
import java.util.stream.IntStream;

/**
 * This class provides a demonstration of the speed-up that can be obtained
 * by using parallel streams rather than sequential streams.
 */
public class ParallelPrimesDemo {

    // A deliberately inefficient way of determining whether an integer
    // n>=2 is prime. We just want to perform a computationally intensive
    // task to demonstrate the benefits of parallelism.
    private static boolean isPrime(int n) {
        for (int i = 2; i < n; i++) {
            if (n % i == 0) {
                return false;
            }
        }
        return true;
    }

    public static void main(String[] args) {
        // Step 1. Create a large array of random integers. We ensure that
        // each integer is greater than or equal to 2, so that it makes
        // sense to ask whether the integer is prime.
        Random random = new Random();
        final int numValues = 2000000;
        final int maxValue = 10000;
        int[] values = new int[numValues];
        for (int i = 0; i < values.length; i++) {
            values[i] = 2 + random.nextInt(maxValue - 2);
        }
```

```java
                // We repeat the remaining steps several times so that we can check
                // if the timing results are reliable.
                int numRepetitions = 5;
                for (int repetition = 0; repetition < numRepetitions; repetition++) {

                    // Step 2. Use a sequential stream to determine how many of the
                    // random integers are prime, and record the time taken.
                    long numPrimes1, numPrimes2;
                    long startTime, endTime;
                    startTime = System.nanoTime();
                    numPrimes1 = IntStream.of(values).sequential()
                                .filter(n -> isPrime(n)).count();
                    endTime = System.nanoTime();
                    long sequentialDuration = (endTime - startTime)
                        / 1000000; // milliseconds

                    // Step 3. The same as the previous step, but with a parallel
                    // stream.
                    startTime = System.nanoTime();
                    numPrimes2 = IntStream.of(values).parallel()
                                .filter(n -> isPrime(n)).count();
                    endTime = System.nanoTime();
                    long parallelDuration = (endTime - startTime)
                        / 1000000; // milliseconds

                    // The answers had better be the same!
                    assert numPrimes1 == numPrimes2;

                    // Step 4. Print the timing results.
                    double speedup = (double) sequentialDuration
                                / parallelDuration;
                    System.out.format(
                    "sequential %dms, parallel %dms, speedup factor %2.2f\n",
                            sequentialDuration, parallelDuration, speedup);
                }
        }
}
```

## 4 Concluding remarks on functional programming

The goal of this chapter was to introduce some of the ideas in functional programming. We focused on the use of functions as first-class objects, meaning that functions can be assigned to local variables, passed as parameters, and returned from other functions. This, however, is just one aspect of functional programming. When writing a purely functional program, another key concept is *immutability*. This refers to the fact that the state of an object is never changed once it is created. Some Java objects have this property. For example, Java `String` objects and `Stream` objects are essentially immutable. But in a purely functional program, everything is immutable. Therefore, the functions have no side effects. This is in stark contrast to Java, where fucntions (i.e. methods) frequently alter the state of an object. For example, a mutator method such as `setColor(Color c)` in the `java.awt.Graphics` class changes the state of the calling `Graphics` object. The side effect is that subsequent invocations of drawing methods like `drawRect()` will draw in a different color. In pure functional programming, there are no side effects. This can make it easier to write and debug code correctly, especially for multithreaded code. To summarize, in this chapter we have studied only one of the important aspect of functional programming: lambda expressions. To fully appreciate the functional paradigm, it is recommended to study a fully-featured functional programming language such as Haskell, Lisp, Scheme, or F#.

# 5 Solutions to example problems

Example problem 1.

```
applyToMinusNine(add5) → -4
applyToMinusNine(multBy3IfPositive) → 0
```

Example problem 2.

```
applyToMinusNine(add5) → -4
applyToMinusNine(multBy3IfPositive) → 0
```

Example problem 3.

```
applyToSeven(lambda potato: potato%4 + potato*potato) → 52
applyToMinusNine(lambda oak: math.factorial(oak+12) * oak) → -54
```

Example problem 4.

```
val5 = 56
val6 = -90
```

Example problem 5.

```
(u,v,w)->Math.sqrt(u*u + v*v + w*w)
```

Example problem 6.

There are many ways to achieve this. Here are some examples:

(i) `Stream<Double> doubleObjects = Stream.of(23.4, 69.7, -25.88, 31.3363);`

(ii) `DoubleStream doublePrimitives = DoubleStream.of(23.4, 69.7, -25.88, 31.3363);`

(iii) Either of the following approaches would work here:
```
IntStream range1 = IntStream.range(100, 201);
IntStream range2 = IntStream.rangeClosed(100, 200);
```

Example problem 7.

```java
Stream<String> gatsby = Files.lines(Paths.get("GreatGatsby.txt"));
long numLines = gatsby.count();
```

Example problem 8.

```java
int sum27to159 = IntStream.rangeClosed(27, 159).sum();
```

Example problem 9.

```java
int sumOdd27to159 = IntStream.rangeClosed(27, 159)
                      .filter(x -> x % 2 == 1)
                      .sum();
```

Example problem 10.

```java
IntStream.rangeClosed(27, 159).forEach(System.out::println);
```

Example problem 11.

```java
Stream<String> info = Files.lines(Paths.get("info.txt"));
info.map(s->s.substring(0, 1)).forEach(System.out::println);
```

Example problem 12.

```java
Stream<String> info = Files.lines(Paths.get("info.txt"));
info.mapToInt(s->s.length()).forEach(System.out::println);
```

Example problem 13.

```java
Stream<String> numbers = Files.lines(Paths.get("numbers.txt"));
BigDecimal sum = numbers
         .map(s -> new BigDecimal(s))
         .reduce(new BigDecimal(0), (tot, val) -> tot.add(val));
System.out.println(sum);
```

Example problem 14.

The following results were obtained on an Intel Core i5-8350U Processor:

```
sequential 2907ms, parallel 634ms, speedup factor 4.59
sequential 2790ms, parallel 639ms, speedup factor 4.37
sequential 2714ms, parallel 651ms, speedup factor 4.17
sequential 2729ms, parallel 651ms, speedup factor 4.19
sequential 2731ms, parallel 646ms, speedup factor 4.23
```

This processor is described by Intel as having 4 cores and 8 threads. One plausible interpretation of the above speedup factors is that each of the four cores was able to perform part of the processing simultaneously, and a small additional boost was provided by the hyperthreading technology on this processor.